\documentclass[11pt,twoside]{article}
\usepackage{asp2010}

\resetcounters

\begin{document}

\title{Natures of clump-origin bulges: similarities to the Milky Way bulge}
\author{Shigeki Inoue$^1$
\affil{$^1$Mullard Space Science Laboratory, University College London, Holmbury St. Mary, Dorking, Surrey, RH5 6NT, United Kingdom}}

\begin{abstract}
Bulges in spiral galaxies have been supposed to be classified into two types: classical bulges or pseudobulges. Classical bulges are thought to form by galactic merger with bursty star formation, whereas pseudobulges are suggested to form by secular evolution. \citet{n:98,n:99} suggested another bulge formation scenario, `clump-origin bulge'. He demonstrated using a numerical simulation that a galactic disc forms clumpy structures in the early stage of disc formation, then the clumps merge into a single bulge at the centre. I perform a high-resolution $N$-body/SPH simulation for the formation of the clump-origin bulge in an isolated galaxy model. I find that the clump-origin bulge resembles pseudobulges in dynamical properties, but this bulge consists of old and metal-rich stars. These natures, old metal-rich population but pseudobulge-like structures, mean that the clump-origin bulge can not be simply classified into classical bulges nor pseudobulges. From these results, I discuss similarities of the clump-origin bulge to the Milky Way (MW) bulge.
\end{abstract}

\section{Introduction}
\label{intro}
\cite{kk:04} has suggested that bulges in spiral galaxies can be classified into \textit{classical bulges} or \textit{pseudobulges}. Classical bulges are thought to form through galactic merger. Pseudobulges are discussed to form through secular evolution caused by non-axisymmetric structures in a galactic disc. \cite{n:98,n:99} demonstrated that clumpy structures form due to gas instability, which could also explain some clumpy galaxies observed in the high-redshift Universe. These galaxies are referred to as clump clusters (chain galaxies). \cite{n:98,n:99} suggested that these clumpy stellar structures fall into the galactic centre by dynamical friction and merge into a single bulge at the galactic centre, a clump-origin bulge. Clump-origin bulges form through `\textit{mergers of the clumps}' in a galactic disc, neither the galactic merger nor the secular evolution. Therefore, properties of clump-origin bulges could be different from those of the conventional ones, classical bulges nor pseudobulges. I perform a similar numerical simulation to \cite{n:98,n:99} using an isolated halo model by a $N$-body/SPH code and study the naive natures of clump-origin bulges in details. 

\section{Results}
Our initial condition follows the spherical model that was used to study the formation of disc galaxies in an isolated environment. I assume an equilibrium system with the NFW profile with a virial mass $M_{\rm vir} = 5.0\times10^{11}~{\rm M_\odot}$. Baryon mass fraction of the system is set to $0.06$. The details of my simulation settings are given in \cite{is:11a,is:11b}.

\begin{figure}
  \begin{minipage}{0.47\hsize}
    \begin{center}
      \includegraphics{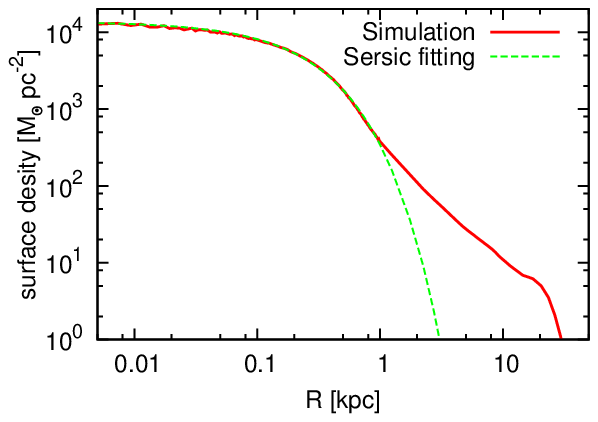}
    \end{center}
  \end{minipage}
  \begin{minipage}{0.4\hsize}
    \begin{center}
      \includegraphics{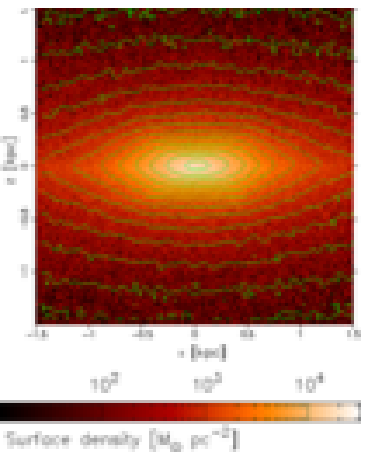}
    \end{center}
  \end{minipage}

  \caption{Stellar surface density profile and a fitting by the S\'ersic profile (left). Stellar surface density maps in central $3\times3$ kpc region from the edge-on view (right).}
  \label{fig:1}       
\end{figure}

In Fig. \ref{fig:1}, I plot the azimathally averaged surface density and density map from the edge-on view. The fitting is given by the S\'ersic index, $n = 1.18$ indicating a nearly exponential density profile. As seen from the central panel, it clearly appears that this bulge is a boxy bulge from the edge-on view. Furthermore, I find that this bulge shows a significant rotation with a value of $V_{\rm max}/\sigma_0\simeq0.9$, the rotation (spin) is \textit{not} negligible in kinematics. These are indicating pseudobulge signatures \cite{kk:04}.

However, I find that the clump-origin bulge consists of stars with an over-solar metallicity \cite{is:11b}. Additionally, this bulge formation scenario, the clump cluster phase, is expected to happen only at the high-redshift. Therefore, the clump-origin bulge consists of old stars. Such old and metal-rich natures are better similar to classical bulges rather than pseudobugles \cite{kk:04}.

The MW bulge is also known to be an ambiguous bulge. The MW bulge shows a nearly exponential profile, an oblate peanut shape (X-shape) and a significant rotation, which are similar to pseudobulges. At the same time, the MW bulge is made of old and metal-rich stars, which are classical bulge signatures \cite{kk:04}. These properties of the MW bulge are consistent with the clump-origin bulge obtained in this study. Such unclassifiable bulges (old pseudobulge) are also observed in some other disc galaxies \cite{ba:87}. My simulation results imply that such old pseudobulges like the MW bulge may be a clump-origin bulge and the MW might use to be a clump cluster.



\end{document}